\documentclass[usegraphicx,useAMS,usenatbib]{mn2e}
\usepackage{url}
\usepackage{lscape}   
\usepackage{multicol} 
\usepackage{verbatim} 
\usepackage{multirow} 
\usepackage{rotating} 
\usepackage{xspace}
\usepackage[usenames, dvipsnames]{color} %Purely there to make coloured text

\def \kms{\mbox{km\,s$^{-1}$}\xspace}
\def \Kkms{\mbox{K\,km\,s$^{-1}$}\xspace}
\def \vlsr{V$_{\rm LSR}$\xspace}
\def \arcsec{{$^{\prime\prime}$}\xspace}
\def \cmcube{\mbox{cm$^{-3}$}\xspace}

\def \trms{\mbox{T$_{\rm RMS}$}\xspace}
\def \vrms{\mbox{$v_{\rm RMS}$}\xspace}
\def \tmb{\mbox{T$_{\rm mb}$}\xspace}
\def \txb{\mbox{T$_{\rm xb}$}\xspace}
\def \ta{\mbox{T$_{\rm A}^{*}$}\xspace}
\def \tsys{\mbox{T$_{\rm SYS}$}\xspace}
\def \nh3{\mbox{NH$_{3}$}\xspace}
\def \nhone{\mbox{NH$_{3}$\,(1,1)}\xspace}
\def \nhtwo{\mbox{NH$_{3}$\,(2,2)}\xspace}
\def \nhthree{\mbox{NH$_{3}$\,(3,3)}\xspace}

\def \msun{\mbox{M$_{\odot}$}\xspace}
\def \miriad{\mbox{\sc Miriad}\xspace}

\def \livedata{\mbox{\sc Livedata}\xspace}
\def \gridzilla{\mbox{\sc Gridzilla}\xspace}
\def \etal{et al.\xspace}

\title[Shocked gas towards HESS\,J1801$-$233]
{Ammonia excitation imaging of shocked gas towards the W28 gamma-ray source HESS\,J1801$-$233}
\author[N. Maxted et al.]
       {Nigel I. Maxted$^{1}$\thanks{E-mail: n.maxted@unsw.edu.au},
       	Phoebe de Wilt$^{2}$,
        Gavin P. Rowell$^{2}$, 
       	Brent P. Nicholas$^{2}$\thanks{New address circa 2012, Cyber and Electronic Warfare Division, Defence Science and Technology Organisation, PO Box 1500, Edinburgh SA 5111, Australia},
        \newauthor Michael. G. Burton$^{1,3}$, 
        Andrew Walsh$^{4}$, 
        Yasuo Fukui$^{5}$, 
        and Akiko Kawamura$^{6}$.\\
$^{1}$School of Physics, University of New South Wales, Sydney, 2052, Australia\\
$^{2}$School of Physical Sciences, Adelaide University, Adelaide, 5005, Australia\\
$^{3}$Armagh Observatory and Planetarium, College Hill, Armagh, BT61 9DG, Northern Ireland, United Kingdom\\
$^{4}$International Centre for Radio Astronomy Research, Curtin University, GPO Box U1987, Perth, Australia\\
$^{5}$Department of Astrophysics, Nagoya University, Furocho, Chikusa-ku, Nagoya, Aichi, 464-8602, Japan\\
$^{6}$National Astronomical Observatory of Japan, Mitaka, Tokyo 181-8588, Japan}

\begin{document}

\date{7th July 2016}%**** ***** **; Received 27 June 2014; in original form \today}

%\pagerange{\pageref{firstpage}--\pageref{lastpage}} \pubyear{2010}

\maketitle

\label{firstpage}

\begin{abstract}
We present 12\,mm Mopra observations of the dense ($>$10$^3$\,cm$^{-3}$) molecular gas towards the north-east (NE) of the W28 supernova remnant (SNR). This cloud is spatially well-matched to the TeV gamma-ray source HESS\,J1801$-$233 and is known to be a SNR-molecular cloud interaction region.
%, which are $\sim$2.5 times less noisy than our previous 12\,mm mapping campaigns, achieving a noise level of $\sigma\sim$0.03\,K.
%The NE cloud is spatially well-matched to the TeV gamma-ray source HESS\,J1801$-$233 and is known to be a SNR-molecular cloud interaction region. 
%Our observations provide \nh3 inversion transition spectra that are $\sim$2.5 times less noisy than our previous 12\,mm mapping campaigns, achieving a noise level of $\sigma\sim$0.03\,K, revealing extended \nh3(6,6) emission. 
%Shock-disruption is evident from \nhone spectra which have unresolvable hyperfine components due to line-width broadening in regions towards the W28 SNR, while strong detections of spatially-extended \nhthree, NH$_3$(4,4) and NH$_3$(6,6) inversion emission towards the cloud strengthen the case for the existence of high temperatures within the cloud. 
Shock-disruption is evident from broad \nhone spectral line-widths in regions towards the W28 SNR, while strong detections of spatially-extended \nhthree, NH$_3$(4,4) and NH$_3$(6,6) inversion emission towards the cloud strengthen the case for the existence of high temperatures within the cloud. 
%Velocity dispersion measurements confirm the conclusion of previous NH$_3$ studies, that the W28 SNR is a source of disruption, and now we have extended the study to include NH$_3$(4,4) emission. 
Velocity dispersion measurements and NH$_3$(n,n)/(1,1) ratio maps, where n=2, 3, 4 and 6, indicate that the source of disruption is from the side of the cloud nearest to the W28 SNR, suggesting that it is the source of cloud-disruption.
%demonstrate a larger flux of higher J/K-level inversion transition emission towards the W28 side of the cloud, both suggest that the W28 SNR is the source of cloud-disruption.} 
Towards part of the cloud, the ratio of ortho to para-NH$_3$ is observed to exceed 2, suggesting gas-phase NH$_3$ enrichment due to NH$_3$ liberation from dust grain mantles. The measured NH$_3$ abundance with respect to H$_2$ is $\sim (1.2 \pm 0.5)\times10^{-9}$, which is not high, as might be expected for a hot, dense molecular cloud enriched by sublimated grain-surface molecules. The results are suggestive of NH$_3$ sublimation and destruction in this molecular cloud, which is likely to be interacting with the W28 SNR shock.
\end{abstract}


\section{Introduction}

W28 is a mature ($>10^{4}$\,yr, \citealt{kaspi1993}) mixed-morphology supernova remnant
(SNR) and a prime example of a region of TeV (10$^{12}$\,eV) gamma-ray excess overlapping with molecular gas \citep{hess:w28}; one indicator of a hadronic production mechanism.
W28 is estimated to be at a distance of 1.2-3.3\,kpc (e.g. \citealt{goudis1976,lozinskaya1981,motogi2010}) and has been detected from radio to gamma-ray energies (e.g. \citealt{dubner2000,rho2002,hess:w28,fermi:w28,agile:w28,Nakamura:2014}).

Of particular interest are the molecular clouds north-east (NE) of
W28. Towards here, molecular emission lines have broad
profiles \citep{arikawa1999,torres2003,reach2005,nicholas2011a,
  Nicholas:2012} and the presence of many 1720\,MHz OH \citep{DeNoyer:1983,frail1994,claussen1997} and 44\,GHz CH$_3$OH \citep{Pihlstrom:2014} masers suggest that the W28 SNR shock is disrupting the clouds. Furthermore, observations targeting the DCO$^+$/HCO$^+$ molecules in the north of these clouds suggest the presence of elevated levels of ionisation consistent with the existence of a nearby source of 0.1-1\,GeV cosmic rays \citep{Vaupre:2014}.

\begin{figure}
\includegraphics[width=\columnwidth]{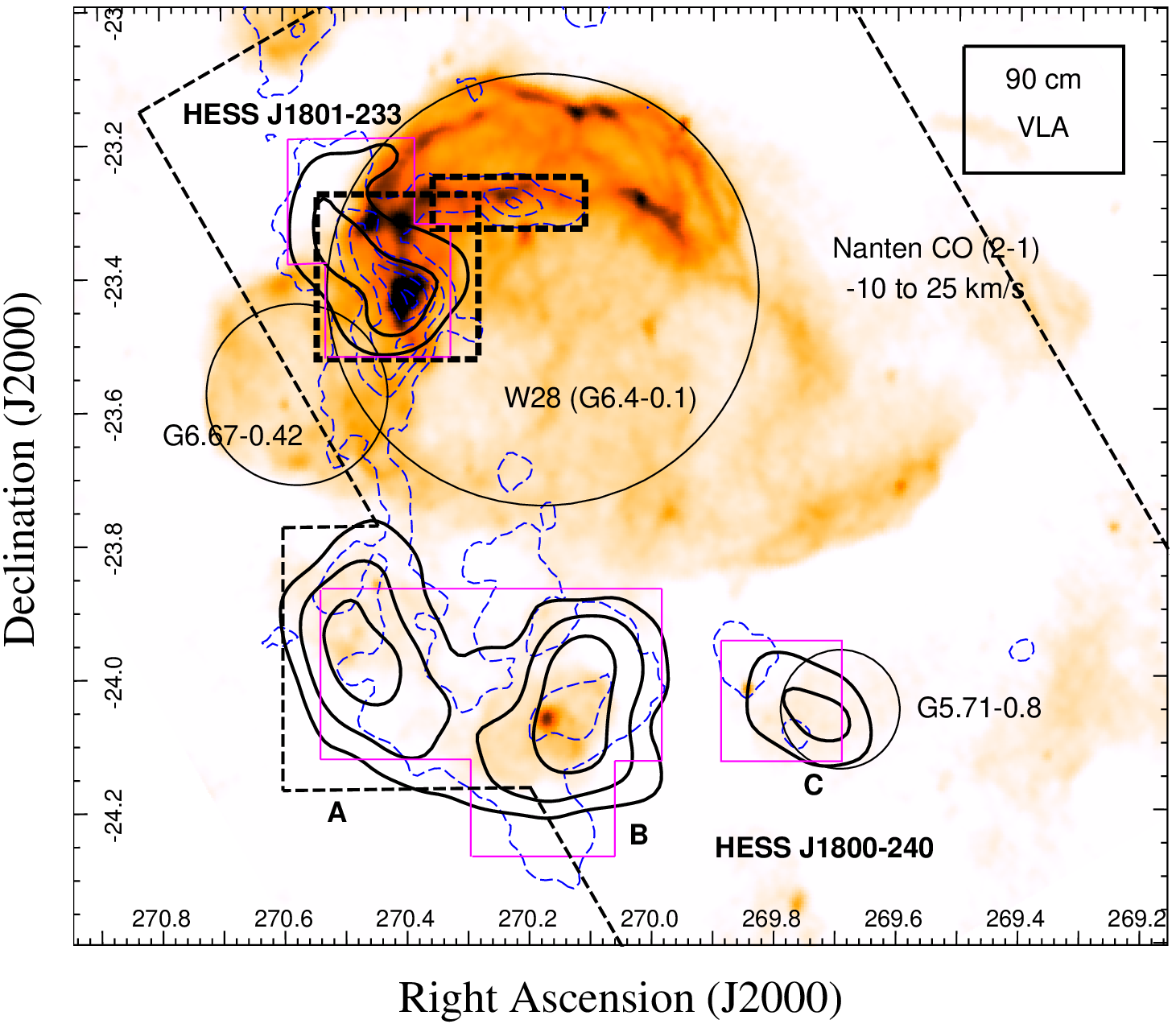}
\caption{An image of 90\,cm continuum emission (linear 0.06-0.55\,Jy\,beam$^{-1}$) observed with the VLA \citep{brogan2006}. TeV gamma-ray emission significance (3-6\,$\sigma$) contours (thick, black) and CO\,(2-1) emission contours (dashed, blue) are overlaid \citep{hess:w28,fukui2008}.The approximate boundaries of catalogued SNRs \citep{brogan2006,yusef2000} are displayed as black thin circles.%, and the positions of 1720\,MHz OH masers are indicated by $+$ \citep{claussen1997,hewitt2009}. 
Regions from previous 12\,mm mapping campaigns (including HOPS data from \citealt{walsh2008}) are enclosed by a thin black dashed boundary, whereas thin, solid magenta boundaries indicated previous 7\,mm mapping regions. Thick dashed black boxes indicate the mapped regions in this work.}
\label{fig:overview}
\end{figure}

% 12mm Last time...
In an attempt to understand the disruption and dynamics of all the
molecular clouds surrounding W28, \citet{nicholas2011a} conducted
broad scale ($\sim 1.5^{\circ}$\,square) observations of the W28 field
in a 12\,mm line survey with $\sim 2^{\prime}$ FWHM resolution.
The dense interiors of the molecular clouds towards the W28 TeV gamma-ray sources, HESS\,J1801$-$233 and HESS\,J1800$-$240 (sub-regions A, B and C), were probed with \nh3 inversion transitions observed at 12\,mm with the Mopra radio telescope. Multiple dense clumps and cores spatially-consistent with both the CO-traced gas, and TeV gamma-ray sources were revealed. Also, the extent to which the W28 SNR has disrupted the dense core of the NE cloud at line of sight velocity $\sim 7$\,kms$^{-1}$ was shown. Strong \nhthree emission, and NH$_3$(6,6) emission suggested this region is warm and turbulent \citep{nicholas2011a}. Modelling of the dense gas in the NE cloud with the MOLLIE radiative transfer software \citep{Keto:1990} suggested that the inner dense cloud component has mass $>1300$\,\msun.
% 7mm last time...
Further observations toward the W28 field were conducted in a 7\,mm
line survey \citep{Nicholas:2012}, which offered superior angular resolution ($\sim 1^{\prime}$ FWHM) relative to 12\,mm observations. The J=1-0 transition of the CS molecule and isotopologues, C$^{34}$S and $^{13}$CS, were used as an independent probe of the dense gas in the region. Simultaneously-observed SiO(1-0) emission exposed the sites of shocks and/or outflows. Both CS(1-0) and SiO(1-0) emission were detected towards the NE cloud, revealing sub-structure in the shocked cloud that lower sensitivity \nh3 observations did not resolve. 
Figure\,\ref{fig:overview} indicates regions which have been mapped in previous molecular emission mapping campaigns towards the W28 SNR field.

% Promote Deep obs..
The broad spectral profiles from all lines detected in the NE cloud
indicate that a kinetic energy of $\sim 10^{48}$\,erg is contained within turbulent gas motions \citep{nicholas2011a,Nicholas:2012}, and it is possible that multiple gas components exist. Detailed \nh3 spectra from across the entire cloud core may help to accurately determine the cloud temperature and density gradients, thus providing better constraints on the total dense cloud mass. Such constraints are important for investigations of the cosmic ray density in a hadronic scenario for gamma-ray emission (e.g. \citealt{Maxted:2013a,Maxted:2013b}), because the measured gamma-ray flux is proportional to both the gas mass and cosmic ray density.

%  conclude this work
To further probe the structure of the dense and disrupted gas towards the NE of W28, the Mopra radio telescope is used to create deeper 12\,mm NH$_3$ inversion transition maps. These observations provide better sensitivity than any previous large scale dense gas studies of the region.% and allow for the first time a pixel-by-pixel estimation of gas parameters via \nh3 inversion transition analysis.

\section{Mopra Observations and Data Reduction}\label{sec:obs}
The observations were performed with the Mopra radio telescope in
April of 2010. We have also included the earlier observations from the
2008 and 2009 seasons \citep{nicholas2011a}, as well as data from the H$_{2}$O Southern Galactic Plane survey \citep{walsh2008} where possible (see Figure\,\ref{fig:overview}). Raw data are available from the Australia Telescope National Facility data archive\footnote{www.atoa.atnf.csiro.au} under the project code M519 and data products are published online. All of these observations have utilised the UNSW Mopra wide-band spectrometer (MOPS) in zoom mode. Mopra is a 22\,m single-dish radio telescope located $\sim$450\,km northwest of Sydney, Australia ($31^\circ
16^\prime 04^{\prime\prime}$\,S, $149^\circ 05^\prime
59^{\prime\prime}$\,E, 866m a.s.l.).
The 12\,mm receiver operates in the 16-27.5\,GHz frequency range. The spectrometer, MOPS, allows an instantaneous 8\,GHz bandwidth. MOPS can record 16 different 137.5\,MHz-wide windows simultaneously when in `zoom'-mode. Each of these 16 windows contains 4096 channels in each of two polarisations.
At 12\,mm this gives MOPS an effective bandwidth of $\sim$1800\,\kms
with a resolution of $\sim$0.4\,\kms. 
Across the whole 12\,mm band, the beam FWHM varies from 2.4$^{\prime}$
(19\,GHz) to 1.7$^{\prime}$ (27\,GHz) \citep{urquhart2010}.

\begin{comment}
Table \ref{tab:lines} lists the lines which MOPS was tuned to receive
in our observations.
\begin{table}
\centering
\caption{Molecular lines and their corresponding rest frequencies
  which MOPS was tuned to receive. The final column indicates whether
  the line was detected in our mapping.\label{tab:lines}}
\normalsize
\begin{tabular}{llc}
\hline
Molecular Line & Frequency & Detected \\
\multicolumn{1}{c}{Name} & \multicolumn{1}{c}{(MHz)} \\
\hline
H69\,$\alpha$    & 19591.1100 & --\\
CH$_{3}$OH - I I & 19967.3960 & --\\
\nh3(8,6)      & 20719.2210 & --\\
\nh3(9,7)      & 20735.4520 & --\\
\nh3(7,5)      & 20804.8300 & --\\
C$_{6}$H         & 20792.8720 & --\\
\nh3(11,9)     & 21070.7390 & --\\
\nh3(4,1)      & 21134.3110 & --\\
H$_{2}$O         & 22235.2530 & Yes\\
\nh3(3,2)      & 22834.1951 & --\\
CH$_{3}$OH - II  & 23121.0240 & --\\
\nh3(2,1)      & 23089.8190 & --\\
H65\,$\alpha$    & 23404.2800 & --\\
CH$_{3}$OH       & 23444.7780 & --\\
\nhone           & 23694.4709 & Yes\\
\nhtwo           & 23722.6336 & Yes\\
\nhthree      & 23870.1269 & Yes\\
\nh3(4,4)      & 24139.4169 & Yes\\
CH$_{3}$OH - I   & 25124.8720 & --\\
\nh3(6,6)      & 25056.0250 & Yes\\
\nh3(7,7)      & 25715.1820 & --\\
\nh3(8,8)      & 26518.9810 & --\\
HC$_{3}$N\,(3-2) & 27294.0780 & --\\
\nh3(9,9)      & 27477.9430 & --\\
CH$_{3}$OH - I   & 27472.5010 & --\\
\hline
\end{tabular}
\end{table}
\end{comment}

We use the MOPS frequency band set-up outlined in \citet{nicholas2011a} to target the TeV source HESS\,J1801$-$233. We completed an additional thirty two passes towards the NE cloud, building on the previous four passes from 2008 and 2009 (making a total of 36 passes). The scanning direction was alternated between right ascension-aligned and declination-aligned to reduce the incidence of scanning artefacts.

\begin{figure*}
\includegraphics[width=0.8\textwidth]{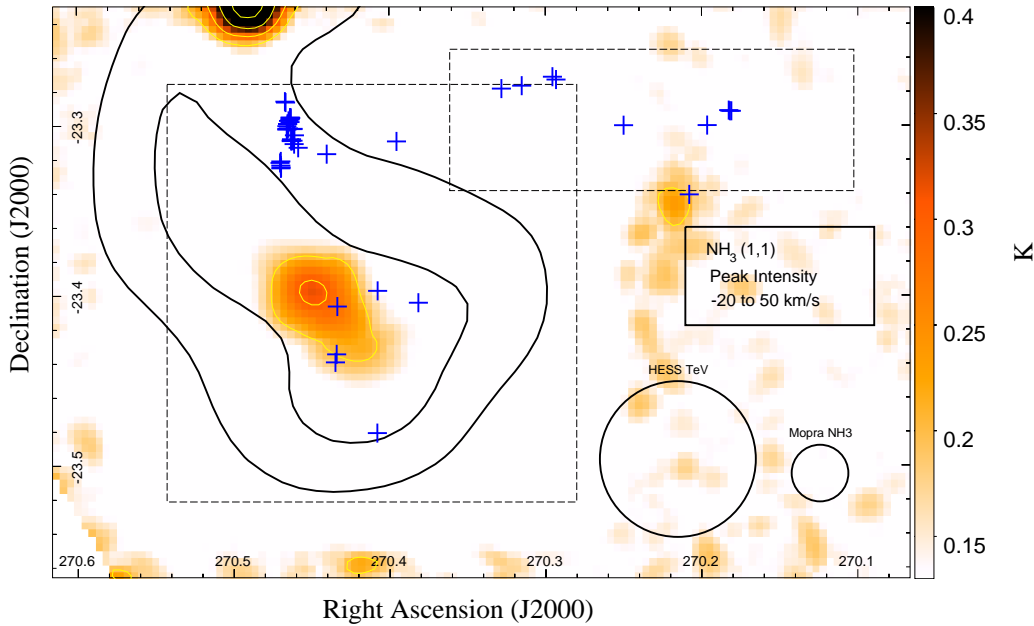}
\caption{Peak \nhone intensity map of the 2 regions (black dashed boxes) in this study. Thick black contours represent the H.E.S.S. TeV statistical significance (4 and 5\,$\sigma$ levels) and yellow contours represent \nhone intensity (T$_A^*$) in increments of 0.1\,K starting from 0.2\,K. OH masers are represented by blue crosses. The exposure time varies over across the displayed field.}
\label{fig:NH3mom-2}
\end{figure*}

In addition to the HESS\,J1801$-$233 field, two new passes were carried out towards a second shocked cloud north of W28 also containing a cluster of 1720\,MHz OH masers (see Figure\,\ref{fig:overview}), but these yielded no new detections (see north-west maser clump in Figure\,\ref{fig:NH3mom-2}), so are not addressed directly in this report.

Data were reduced using the standard ATNF packages \livedata, \gridzilla \citep{Karma} and \miriad \citep{Miriad}. For mapping data, \livedata was used to perform a bandpass calibration for each row, using the preceding off-scan as a reference, and apply a linear fit to the baseline. \gridzilla re-gridded and combined all data from all mapping scans into a single data cube, with pixels ($\Delta\,x,\Delta\,y,\Delta\,z$)=(15\arcsec, 15\arcsec, 0.4\,\kms). The mapping data were also {\tsys}-weighted, smoothed with a Gaussian of FWHM equal to the Mopra beam (2$^\prime$) and a cut-off radius of 5$^\prime$. 

The antenna temperature, \ta, (corrected for atmospheric attenuation and rearward loss) was converted to the main beam brightness temperature, \tmb = \ta /$\eta_{\rm xb}$ where $\eta_{\rm xb}$ is the Mopra extended beam efficiency, which ranged from 0.68 to 0.74 for the NH$_3$ bands \citep{urquhart2010}.

Across the bandpass, the standard deviation in \txb achieved from these mapping observations is $\sim$0.03\,K\,channel$^{-1}$.

%IMAGE PREPARATION
Images of velocity-integrated intensity, position velocity (PV)
and velocity dispersion, \vrms, were produced using \miriad software.
% int. inten
%%%%contours
In integrated emission images, minimum contour levels were set based on the integrated significance of the emission, which was determined on an image by image basis. 

%The integrated \trms of each image was determined by creating $0^{\rm th}$ moment maps in a velocity space either side of the region of interest, ensuring the same number of channels are used. These additional moment maps were used to create pixel distribution histograms, which had Gaussian functions fitted to find the integrated emission \trms.  

% p.v. plot
%Position-Velocity (PV) plots were created by re-ordering the data cubes axes and smoothing the velocity axis with a Hanning function (width $\sim2$\,\kms) to improve image quality. The PV plots only show the peak pixel (\miriad moment -2)
%along the declination axis for illustrative purposes. RA vs \vlsr
%PV plots are displayed in Figure\,\ref{fig:pvplots123}.

% vel. disp
Velocity dispersion (\miriad moment 2), \vrms, maps were calculated
for pixels above a reasonable threshold ($\sim3.5$\,\trms) using the
same method as \citep[Section\,4.3]{nicholas2011a}.

%%%%%%%%%%%%%%%%%%%%%%%%%%%%%%%%%%%%%%%%%%%%%%%%%%%%%%%%%%%%%%%%%%%%%%%%%%%%%%%

\section{Analysis and Results Overview}\label{sec:overview}
The primary targets of the survey were the NH$_3$(1,1) to (4,4), (6,6) and (9,9) transitions, with the goal of determining the location of hot, dense molecular gas. The spectrometer (MOPS) was also tuned to receive the NH$_3$(2-1), (3-2), (4-1), (7-5), (8-6), (9-7) and (11-9) transitions (see \citealt{Nicholas:2012} for line frequencies). Previously, the \nhone, (2,2), (3,3) and (6,6) transitions were detected from the NE cloud. This recent survey has recorded the NH$_3$(4,4) inversion transition, in addition to these. We note that the (5,5) transition was not included in the spectrometer sampling range.

In addition to the NH$_3$ transitions, a 22\,GHz H$_{2}$O maser was detected, but other spectral lines included within the MOPS bandpass, including class II CH$_{3}$OH masers, HC$_{3}$N(3-2) and the recombination lines H65\,$\alpha$ and H69\,$\alpha$, were not detected

\begin{figure*}
\centering
\includegraphics[width=0.99\textwidth]{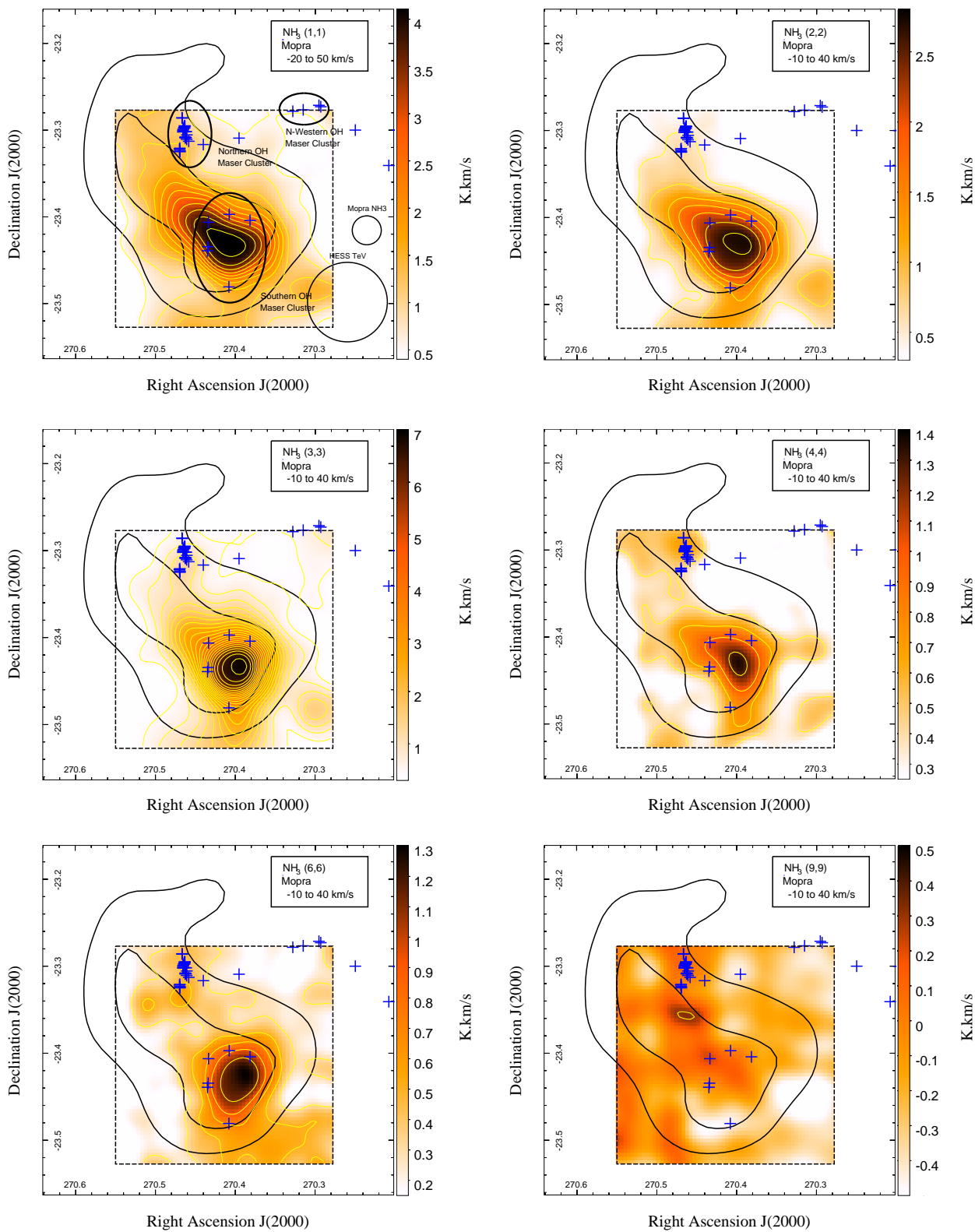}
\caption{Integrated \nh3 emission intensity images (\tmb) towards the NE cloud. HESS TeV gamma-ray emission is indicated by thick, black contours (4 and 5\,$\sigma$ levels). The black dashed box illustrates the region of deep mapping (this campaign). The blue crosses ($+$) indicate the positions of the 1720\,MHz OH masers \citep{claussen1997}. Yellow contours are used to show significant emission from each map. The minimum contour level on all images is 3\,$\sigma$ and increments are in +2\,$\sigma$ levels. The minimum contour levels are 0.51\,\Kkms for \nhone, 0.51\,\Kkms for \nhtwo, 0.45\,\Kkms for \nhthree, 0.42\,\Kkms for NH$_3$(4,4) and 0.3\,\Kkms for NH$_3$(6,6). The integration velocity-range is larger for NH$_3$(1,1) emission to encompass prominent satellite emission lines. All images have been corrected for beam efficiency.}
\label{fig:integrated_multipanel}
\end{figure*}

% discussion of all integrated emissions..
Integrated intensity maps of detected \nh3 transitions are presented
in Figure\,\ref{fig:integrated_multipanel}. The peak of the \nh3
velocity-integrated emission is positionally consistent for the five detected inversion transitions. 
The NH$_3$ features highlighted in \citet{nicholas2011a} are consistent with those in these new images, but now, our deeper mapping reveals weaker features which were not previously seen (at a significant level).

\begin{figure}
\centering
\includegraphics[width=1.05\columnwidth]{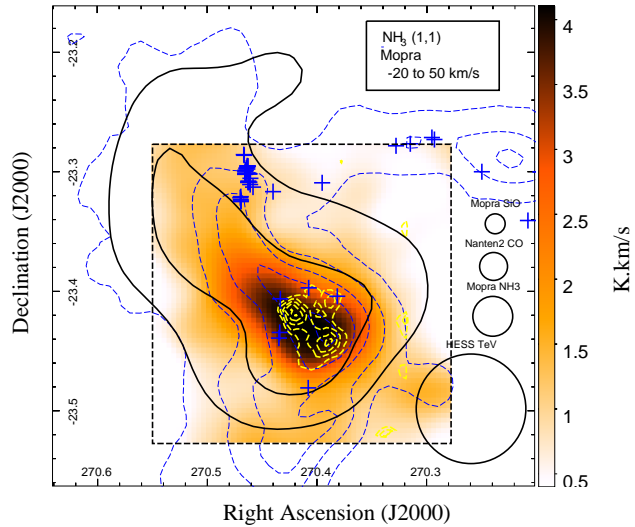}
\caption{\nhone integrated intensity emission image (-20 to 50\,\kms) with H.E.S.S. TeV significance contours (thick solid black, 4 and 5\,$\sigma$ levels), SiO\,(1-0) integrated emission contours (yellow dashed, $\sim$2 to 5\,$\sigma$ levels) from \citealt{Nicholas:2012} and Nanten2 CO(2-1) integrated emission contours (blue dashed). Positions of 1720\,MHz OH masers from \citet{frail1994,claussen1997} are indicated by the blue crosses ($+$).}
\label{fig:comparison1}
\end{figure}

We detect a larger dense component of the NE cloud ($270.45^{\circ},-23.4^{\circ}$), including an extension
of the dense gas protruding south in all detected \nh3 transitions,
following the general distribution of the gas seen by the
Nanten telescope in CO(2-1) emission \citep{nicholas2011a}, as seen in Figure\,\ref{fig:comparison1} \citep{fukui2008}. This \nh3 emission region is also coincident with CS(1-0) emission seen by \citet{Nicholas:2012} and it follows that a significant proportion of the NE cloud is composed of dense, $\sim 10^{4}$\,\cmcube gas or higher (i.e. similar to the critical density for NH$_3$(1,1) emission). %The distribution of the CO(2-1), CS(1-0) and \nhone emission across the NE cloud is displayed in Figure\,\ref{fig:comparison}. From this image, its clear that a significant proportion of the NE cloud is comprised of dense, $n> 3\times10^{3}$\,\cmcube gas (i.e. similar to the critical density for NH$_3$(1,1) emission).

\begin{figure*}
%\flushleft
\centering
\includegraphics[width=0.83\textwidth]{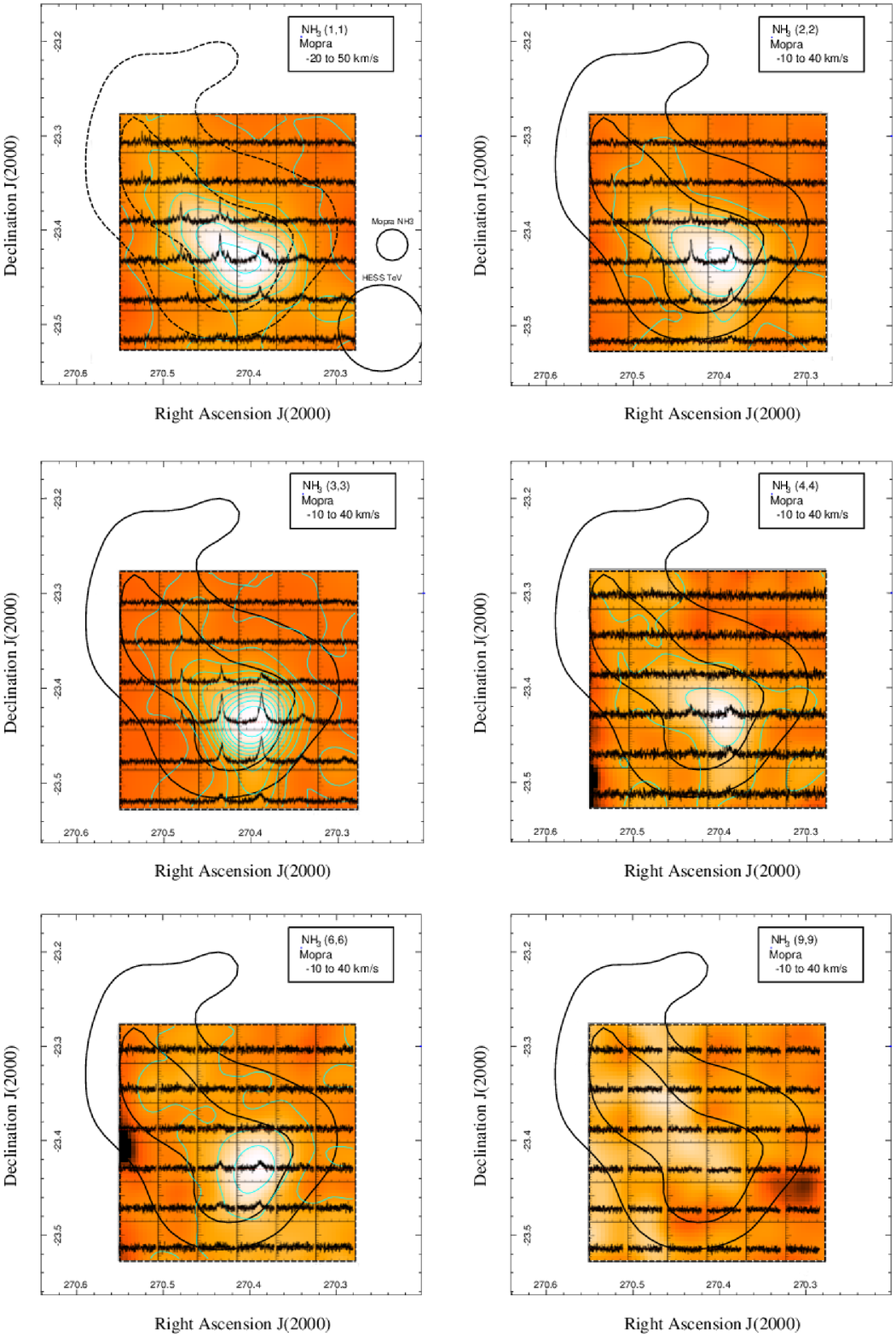}
\caption{Postage stamp plot of the \nh3 emission lines towards the NE cloud of W28. The colour integrated-intensity images represent the same data as in Figure\,\ref{fig:integrated_multipanel}, but with modified colour scale and fewer contour levels (cyan). The average molecular line profile for all pixels within that grid box is provided. The spectral grid limits are the same for all 36 spectra within each emission line image. The velocity axes limits are $-$50 to 50\,\kms. A red dashed line indicates the zero line. For all spectral panels the ordinate lower limit is $-$0.1\,K, whereas the ordinate upper limits are 0.3, 0.3, 0.4, 0.2, 0.2, 0.2\,K for \nhone, \nhtwo, \nhthree, NH$_3$(4,4), NH$_3$(6,6) and NH$_3$(9,9), respectively. HESS TeV gamma-ray emission is indicated by thick, black contours (4 and 5\,$\sigma$ levels).}
\label{fig:postage_plot}
\end{figure*}

An additional dense clump separated from the main \nh3 cloud is detected at an integrated intensity equivalent to a 3 to 4\,$\sigma$ level ($270.30^{\circ},-23.48^{\circ}$) in the \nhone to (4,4) inversion transitions (Figure~\ref{fig:integrated_multipanel}). 
It lies to the south west, outside the 3\,$\sigma$ TeV contour, at a different velocity (\vlsr $\sim$15\,\kms) to that of the majority of the dense gas mass in the region (\vlsr$\sim$7\,\kms), as evidenced in Figure\,\ref{fig:postage_plot}.

`Postage stamp plots' for the detected \nh3 lines are presented in Figure\,\ref{fig:postage_plot}. In this figure, the background integrated intensity images are the same as presented in Figure\,\ref{fig:integrated_multipanel}, but with a modified colour scale (now white displays stronger emission) and fewer contour levels (now cyan). The mapped area is divided into a 6$\times$6 grid and the average line profile from all pixels within that grid box is displayed. These postage stamp plots display additional trends for the \nh3 line emission across the dense cloud component. Generally, the line profiles are broader towards the W28 side of cloud (south-west side of Figure\,\ref{fig:postage_plot}). Here, the line profiles indicate the shocked cloud structure with broad line widths (FWHM $>$ 10\,\kms) and blending of the \nhone satellite components. Additionally, a strong detection of \nhthree ($270.39^{\circ},-23.43^{\circ}$), with a peak intensity $\sim2\times$ larger than the \nhone line, and peaks in the NH$_3$(4,4) and (6,6) lines are seen towards the W28 side (south-west) of the cloud. Further west, towards W28, the line profiles are broad and weak, indicating that a shock may be coming from this direction. Towards the NE of the dense cloud, the line profiles are more `typical' of a cold quiescent cloud. The characteristic satellite lines of \nhone are resolved and the peak intensity of the emission decreases in the higher-excited (2,2) and (3,3) states of NH$_3$.

%A detection of 
The distribution of the \nhone and SiO\,(1-0) emission from the NE cloud is displayed in Figure\,\ref{fig:comparison1}, revealing the position and extent of the shocked and disrupted gas in the dense cloud component. In addition to intense SiO(1-0) emission coincident with the \nhone peak, SiO(1-0) was detected towards the northern OH maser clusters \citep[see][for details]{Nicholas:2012}, coincident with the boundary between the shocked and quiescent gas, as indicated by CO emissions \citep{arikawa1999}. This subregion also corresponds to a brightening of the X-ray shell in the 1-10\,keV energy band \citep{ueno2003,Nicholas:2012}. As can be seen from Figures \ref{fig:integrated_multipanel} and \ref{fig:postage_plot}, emission from \nhone to (4,4) is seen here, and the broadness (6-13\,\kms) of the \nhthree and (4,4) spectral lines indicate the presence of gas with high temperature and disruption, in agreement with indications from the previous shocked gas tracers (SiO, OH).

\subsection{Velocity Dispersion}
\label{ssec:veldisp}
Previously, \citet{nicholas2011a} used a velocity dispersion profile map to illustrate that the western side of NE cloud was experiencing greater disruption than the east. Additionally, the greatest level of dispersion, in both physical area and magnitude, was seen in the \nhthree line.

\begin{figure*}
\includegraphics[width=0.99\textwidth]{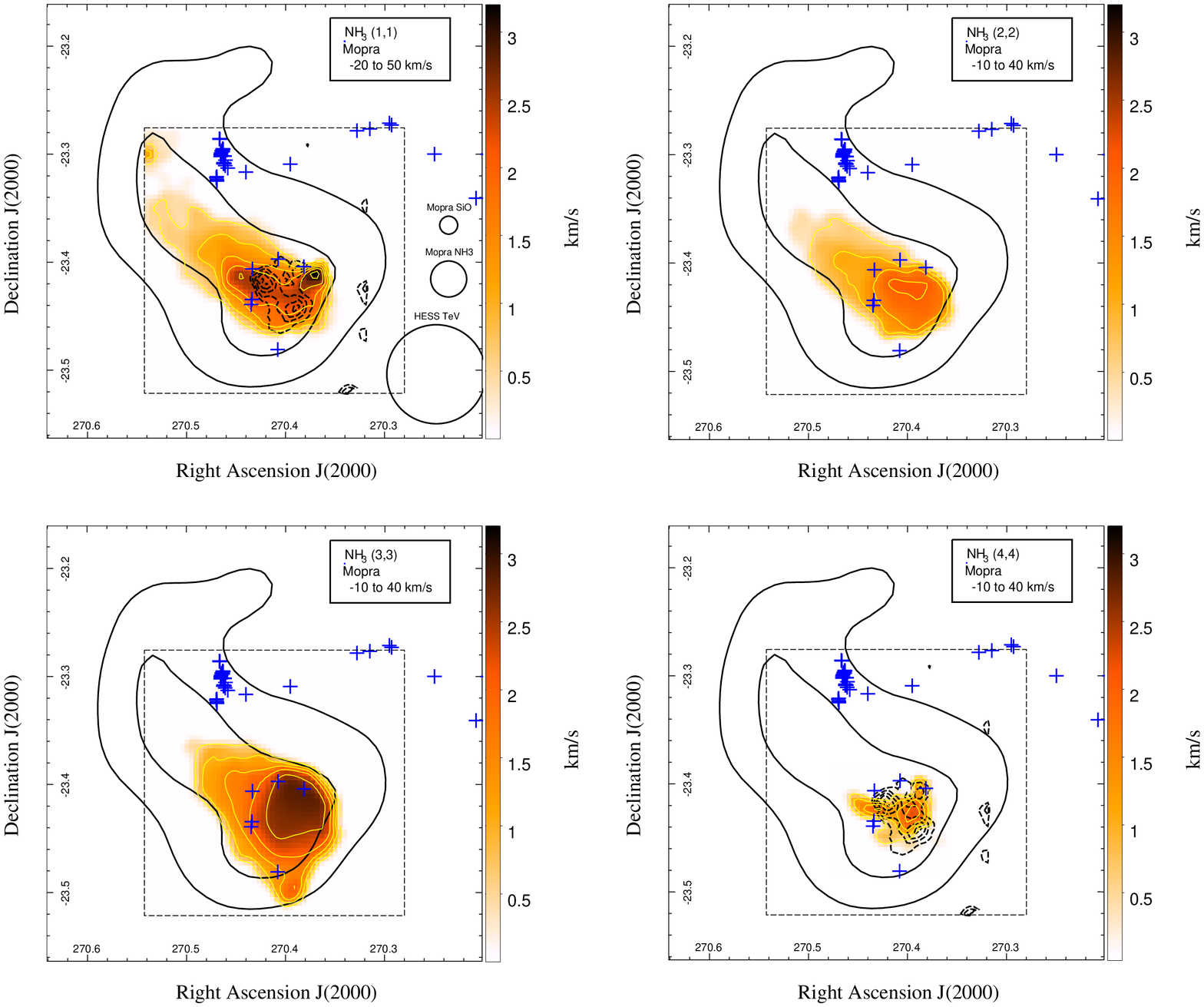} 
\caption{Velocity dispersion, \vrms (km\,s$^{-1}$), maps for the NE cloud. These images are the updated equivalents from \citet[Figure\,9]{nicholas2011a}. In all panels thick black contours are the H.E.S.S. TeV statistical significance (4 and 5\,$\sigma$ levels) and OH masers are represented by blue crosses. The \nhone, \nhtwo, \nhthree and NH$_3$(4,4) dispersions are calculated for pixels with $\tmb\geq3.5$\,$\sigma$ ($\sim$0.11\,K) within a \vlsr range of 5 to 15\,\kms. In 2 images, black dashed contours represent SiO\,(1-0) emission integrated between \vlsr$=-$5 and 20\,km\,s$^{-1}$ ($\sim$3 to 5\,$\sigma$ levels, see \citealt{Nicholas:2012}). We note that the largest region of dispersion is apparent in the \nhthree line.}
\label{fig:veldispmultipanel}
\end{figure*}

Figure\,\ref{fig:veldispmultipanel} presents four velocity dispersion (intensity-weighted FWHM, see \emph{moment\,2} in \citealt{Miriad}) images generated using our sensitive \nh3 transition data. This series of images has similar features to those in \citet{nicholas2011a}, although one new structure is observed. The deeper \nhone data now reveals two peaks or a long finger of broad gas in the dispersion map (compared to only one in \citealt{nicholas2011a}). One peak is towards the centre-east of the cloud core ([$\alpha$,$\delta$]$\sim$[270.44,-23.42]), which was previously observed, whereas a second, slightly stronger peak is detected towards the western side of the cloud([$\alpha$,$\delta$]$\sim$[270.37,-23.42]). The former peak in (1,1) velocity dispersion also appears to correspond to an eastern extension in the (4,4) dispersion image. %In fact, this eastern extension is consistent with a similar extension seen in column density and mass maps, indicating that the densest regions may correspond to the most turbulent regions.

We note that \citet{Nicholas:2012} showed shock-tracing SiO(1-0) emission towards the western side of the cloud, but this new NH$_3$(1,1) velocity dispersion peak lies even further west, as illustrated in Figure\,\ref{fig:veldispmultipanel}. Furthermore, the peak of NH$_3$(4,4) velocity dispersion appears spatially better-matched to the SiO(1-0) emission (with a FWHM a factor 2 less than that of NH3(1,1) emission), suggesting that this is the most disturbed part of the cloud. %and the region of peak column density/mass, highlighting the most disturbed part of the cloud. 
Six 1720\,MHz OH masers lie around the periphery of this region, suggesting that conditions conducive to the generation of 1720\,MHz OH masers directed towards the Earth are not present in the densest, most-energetic region of the north-east cloud, but are instead offset from the position of peak NH$_3$(4,4) emission.% column density. 

\begin{figure*}
\includegraphics[width=0.99\textwidth]{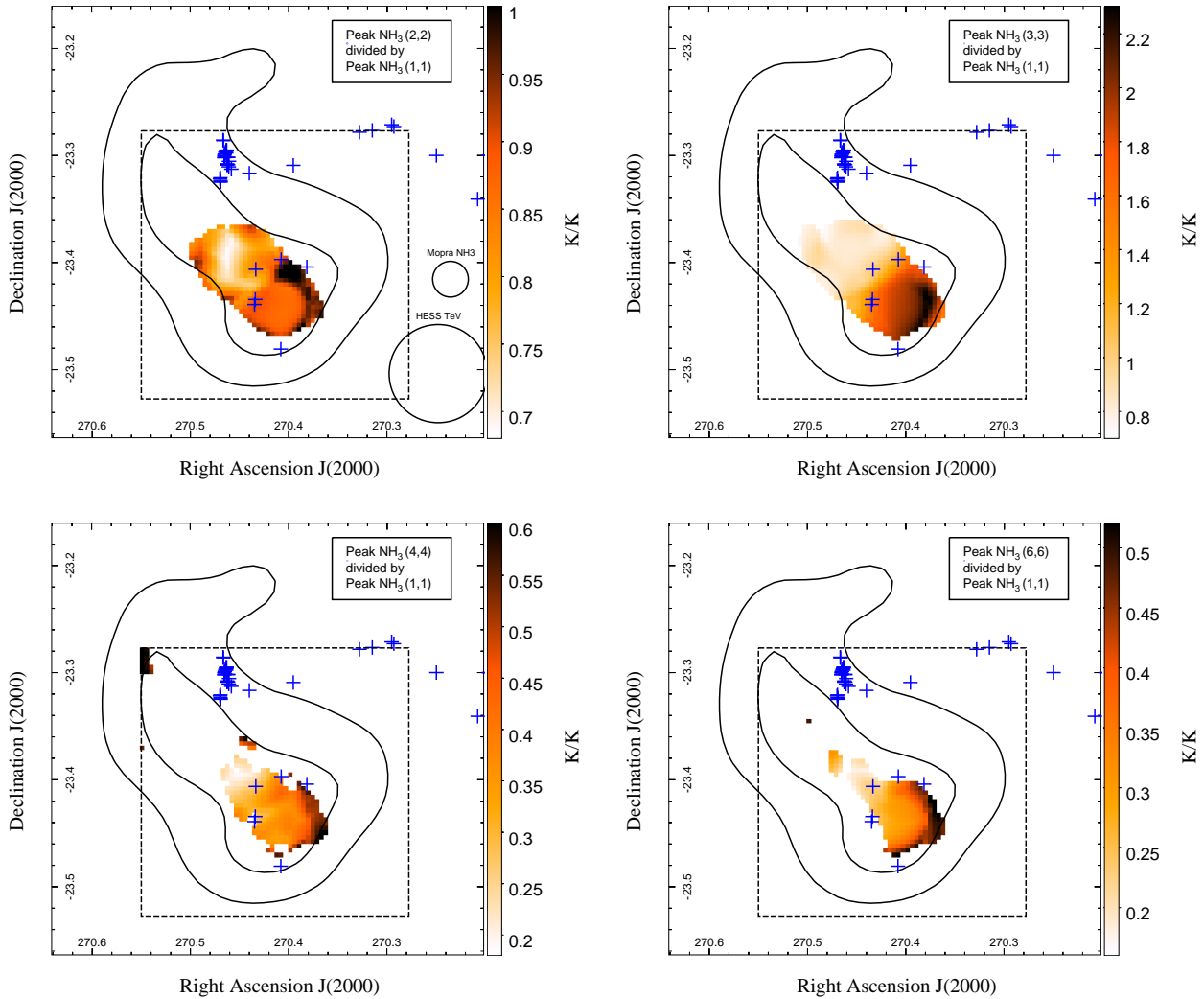} 
\caption{Images of the peak intensity of NH$_3$(n,n) emission divided by the peak intensity of NH$_3$(1,1) emission, where n=2,3,4 and 6. All non-zero pixels had both the constituent peak intensity pixel values exceed a threshold chosen by eye - 0.15, 0.15, 0.13, 0.08 and 0.07\,K for the (1,1), (2,2), (3,3), (4,4) and (6,6) transitions, respectively. OH masers are represented by blue crosses.}
\label{fig:Ratios}
\end{figure*}

Another indication of a disruption occurring from the south-west side of the cloud can be seen in images of peak intensity ratios in Figure\,\ref{fig:Ratios}. The raw values of these images are a product of several inter-related parameters, hence are difficult to directly interpret. The gradients in these images indicate a general trend towards a greater intensity of higher-energy inversion transitions towards the south-west compared to NH$_3$(1,1). This trend is particularly prominent for the NH$_3$(3,3) inversion line, unclear from the NH$_3$(6,6) and less prominent for the NH$_3$(2,2) and (4,4) lines. On examining the NH$_3$(3,3)/NH$_3$(1,1) gradient, it is clear that a temperature gradient is present, although we can't immediately discount the possibility of effects caused by differences in ortho and para-NH$_3$ abundances in a shocked/energetic region (e.g. \citealt{Umemoto:1999}). This issue is considered in Section\,\ref{ssec:Params}.
%The ortho-para NH$_3$ ratio towards gamma-ray sources is a matter currently under investigation \citep{deWilt:2015}. 
Certainly the NH$_3$(3,3) emission does exhibit another unique feature - a southern lobe present in the (3,3) dispersion image (Figure\,\ref{fig:veldispmultipanel}), but not in the (1,1) or (2,2) images. This extension is also seen in CO(3-2) emission \citep{arikawa1999}, so a hot component may be seen to extend towards this same location.

% pixel by pixel
\subsection{LTE Parameter Calculation Prescription}\label{ssec:pixbypix}
The greater sensitivity achieved from deep 12\,mm mapping has allowed us to parametrise the \nh3 satellite lines on a pixel-by-pixel (PbP) basis. An \nh3 analysis could thus be performed on arcminute scales. This includes the calculation of \nhone main line optical depth \citep[Equation 2]{barrett1977}, \nh3 energetic state column densities \citep[Equation\,9]{goldsmith_langer1999}, the rotation temperature (via rotation diagrams) and the total \nh3 column density. A summary of the formulae employed can be found in section B1 of \citet{Maxted:2012}, but the method presented here is adjusted to account for higher temperatures.

% How it works
Our PbP procedure considered each pixel in the \nhone, (2,2), (3,3), (4,4) and (6,6) data cubes separately, and does not attempt to use low signal-noise NH$_3$(9,9) emission. Five Gaussian functions were fitted to the 5 satellite components of \nhone emission, and pixels that had a peak main line intensity $\leq0.13$\,K ($\sim 4$\,\trms) were discarded (set equal to zero). This threshold was decided after the examination of a sample of spectral fits, ensuring that only high-quality spectral parameters were used in our analyses. The optical depth was calculated, and the \nhtwo spectra were fit by single Gaussian functions (satellite lines were generally not resolved for \nhtwo emission). The \nhone, (2,2), (3,3), (4,4) and (6,6) column densities \citep{goldsmith_langer1999} were then calculated for all pixels with an integrated emission threshold above 0.5\,K.km.s$^{-1}$ ($\sim$3\,$\sigma$). Emission from energetic states (J,K)$=$(3,3) and above were assumed to be optically thin.

\begin{figure}
\centering
\includegraphics[width=0.45\textwidth]{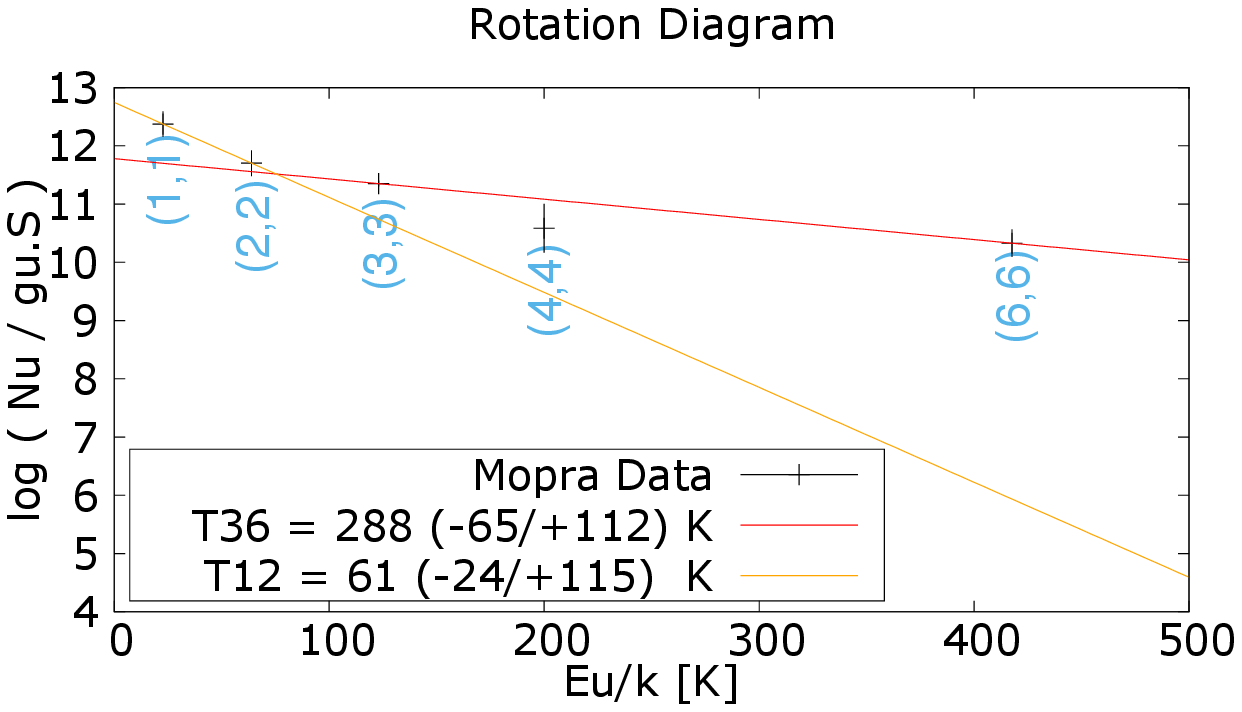}
\caption{An example of a rotational diagram of NH$_3$ inversion transition data towards a pixel representative of the regions with detectable NH$_3$ (1,1), (2,2), (3,3), (4,4) and (5,5) lines. The y-axis is the column density divided by the statistical weight ($g_u$ is rotational degeneracy and $S$ is the spin degeneracy), and the x-axis is the measured NH$_3$ inversion transition energy in kelvin. Two best-fit linear lines are displayed, representing the cold component LTE rotational temperature (orange) and the hot component LTE rotational temperature (red), respectively.}
\label{fig:RotDiag}
\end{figure}

Rotation temperatures $T_{12}$ and $T_{36}$ were calculated using a line of best fit for degeneracy-normalised column density versus transition temperature (e.g. \citealt{Umemoto:1999}), where the subscripts refer to the rotational quantum numbers (J) of the inversion transitions used to derive the temperature. 
%Where the signal-noise level was sufficient (see paragraphs above), rotation temperatures $T_{12}$ and $T_{36}$ were calculated, where the subscripts refer to the rotational quantum numbers (J) of the transitions used to derive the temperature. 
Figure\,\ref{fig:RotDiag} illustrates the difference between the two rotational temperatures attributable to a pixel that is representative of the region with detections of all measured NH$_3$ lines. We attribute this to the existence of `cold' and `hot' components, with the (1,1) and (2,2) transitions being dominated by the more extensive cold component. The region where NH$_3$(6,6), (4,4) and (3,3) emission was observed was treated as a hot component and the ortho-para-ratio was calculated by finding the ratio between the observed NH$_3$(4,4) column density and that inferred from a trend line between the points corresponding to the (3,3) and (6,6) on the rotational diagram. In doing this, we interpolate that $T_{45} = T_{36}$, which is a valid assumption if all the NH$_3$(3,3)-(6,6) emission originates from the same region. Cold and hot component column densities were calculated using the NH$_3$ partition function \citep[see][]{nicholas2011a} assuming temperatures of $T_{12}$ and $T_{36}$, respectively. The calculated OPR was applied in the hot component analysis, but was assumed to be unity for the cold component where the calculation of an OPR was not possible.

The final results were assembled into 2D arrays of optical depths, temperatures and column densities, which were converted into fits files. Fits-file `header' information was copied from the input \nhone fits cube to recreate the axes of the output fits files. The resultant parameter maps are displayed in Figures\,\ref{fig:OptDepth} and \ref{fig:ColDens}.

\subsection{Gas Parameters towards HESS\,J1801$-$233}\label{ssec:Params}
\begin{figure*}
\includegraphics[width=0.99\textwidth]{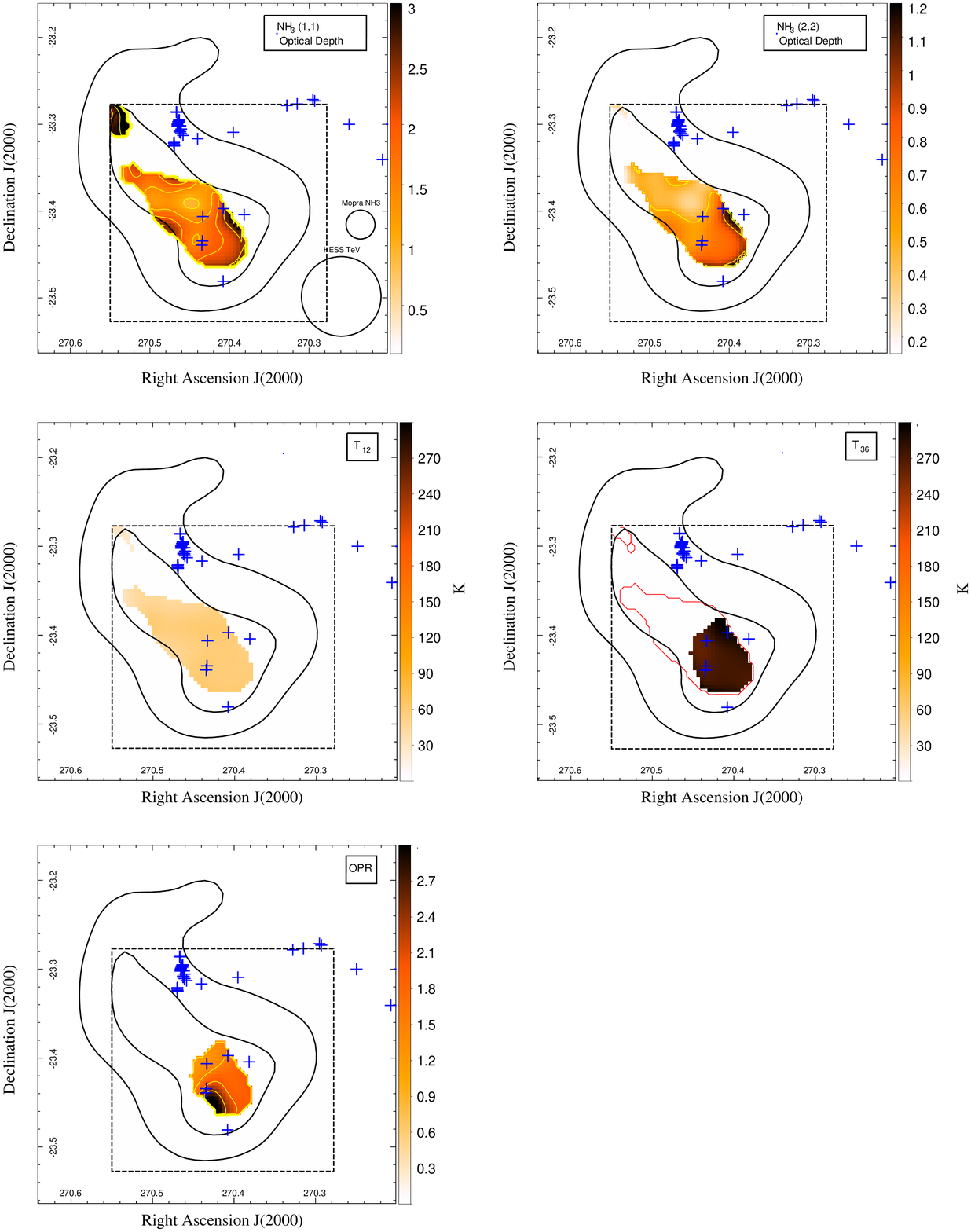} \caption{Images of NH$_3$(1,1) and (2,2) optical depth (top left and right, respectively), the NH$_3$(1,1)/NH$_3$(2,2) rotational temperature, T$_{12}$(middle left), NH$_3$(3,3)/NH$_3$(6,6) rotational temperature, T$_{36}$(middle right), and the ortho-para-NH$_3$ ratio, the OPR (bottom right). Optical depth images have contours in increments of 0.5 from 0. The ortho-para ratio map (OPR, bottom, right) has contours of 1 to 2.5 in increments of 0.5. OH masers are represented by blue (or white, where necessary) crosses in all maps. Maps of the error associated with all of these values are displayed in Figure\,\ref{fig:Errors}.}
\label{fig:OptDepth}
\end{figure*}

\begin{figure*}
\includegraphics[width=0.99\textwidth]{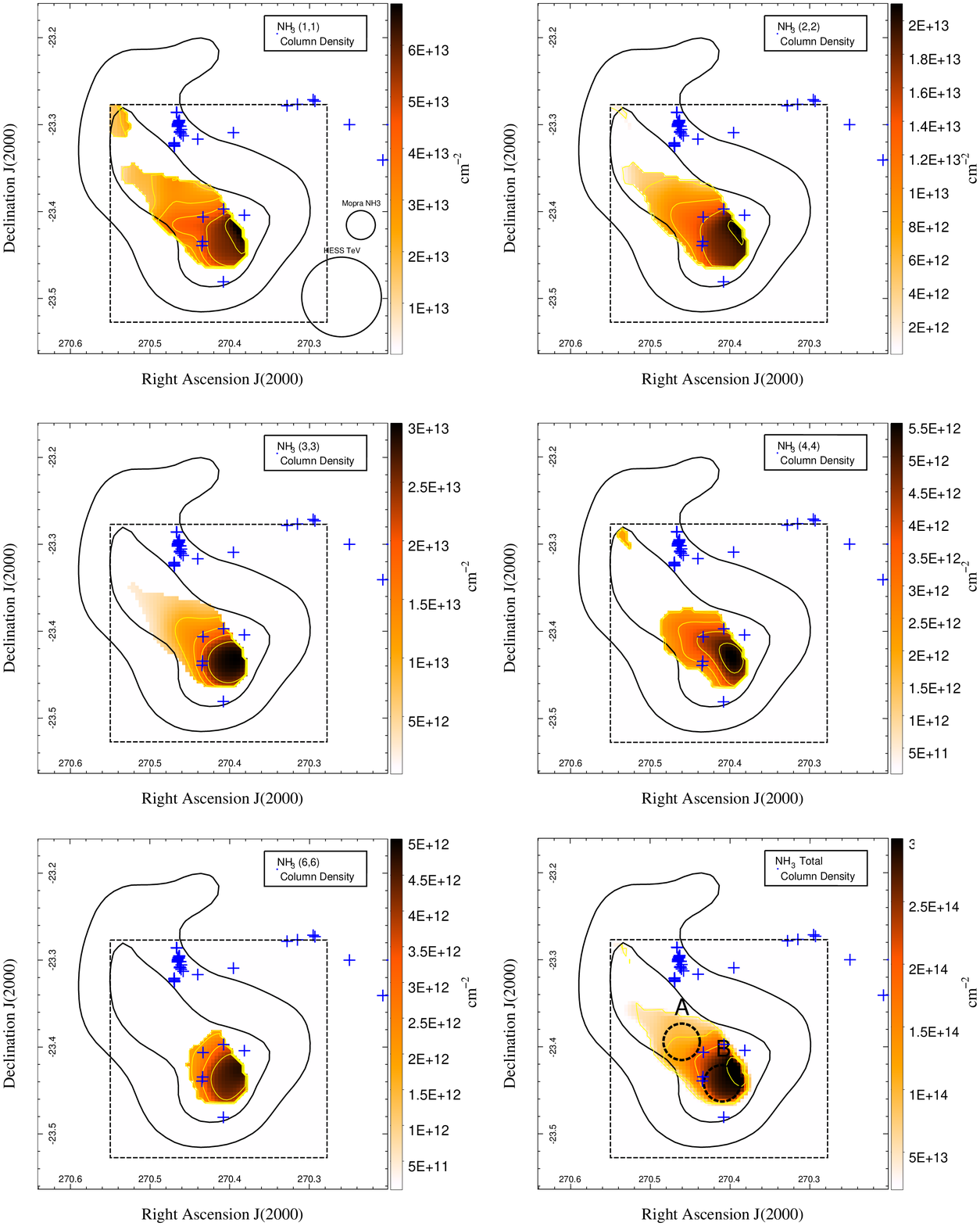} 
\caption{Column densities of the (1,1), (2,2), (3,3), (4,4), (6,6) NH$_3$ states and the total NH$_3$ column density. The total NH$_3$ column density image has been additionally smoothed by a 2D gaussian function with width equal to the Mopra NH$_3$(1,1) beam width, and contains 2 circular regions which were exploited for non-LTE cross-checks (see Section\,\ref{sec:ColdensTemp}). OH masers are represented by blue crosses. NH$_3$(1,1) and total column density error maps are displayed in Figure\,\ref{fig:Errors} and \ref{fig:Error2}, respectively.}
\label{fig:ColDens}
\end{figure*}

Five images are shown in Figure\,\ref{fig:OptDepth}: \nhone main line optical depth, \nhtwo optical depth, the NH$_3$(3,3)/NH$_3$(6,6) rotational temperature and the estimated ortho to para-NH$_3$ ratio (OPR). Figure\,\ref{fig:ColDens} displays the column densities of the (1,1), (2,2), (3,3), (4,4), (6,6) NH$_3$ states, and the total hot + cold component NH$_3$ column density calculated from the method outlined in Section\,\ref{ssec:pixbypix}. In all images, a lower threshold was imposed to ensure the quality of results (see Section\,\ref{ssec:pixbypix}), thus pixels with a value of zero do not represent a value of zero but rather an undefined value.

%optical depth
\subsubsection{Optical Depth}
The NH$_3$(1,1) optical depth map (Figure\,\ref{fig:OptDepth}) reveals an optically thick ($\tau\sim$0.9-3) region corresponding to the peak of \nhone emission (Figure\,\ref{fig:integrated_multipanel}). Some variation is observed on a scale larger than the Mopra NH$_3$(1,1) beam and this may reflect internal clumpiness in the dense gas. At declination$\sim\,-23.38 ^{\circ}$, a region of NH$_3$(1,1) optical depth, $\tau\sim$1, transitions into a region of NH$_3$(1,1) optical depth, $\tau\sim$2, towards the southern part of the cloud at declination$\sim -23.45^{\circ}$. Towards the edges are some regions with optical depths which tend towards $\tau\sim$3. These features are smaller than the NH$_3$(1,1) beam and possibly artefacts introduced by the gaussian function fitting process, when the main emission line intensity becomes more comparable to noise. Towards the western side of the dense cloud (as seen in the integrated NH3 (1,1) emission image, Figure\,\ref{fig:integrated_multipanel}), where NH3 (1,1) profiles become broad and the satellite lines blend, optical depths could not be calculated as the sensitivity was not sufficient.

We note the existence of a relatively optically thick (optical depth $\tau\sim$3$\pm$1) gas component in the north-east corner of the map (Figure\,\ref{fig:OptDepth}, top left). Upon examination of the corresponding \nhone spectra (see Figure\,\ref{fig:postage_plot}), this feature appears to be from a real emission line, but we cannot rule out the high optical depth value being an artefact of the analysis. The flux falls below the threshold limit around this region, so the extent of the feature cannot be determined. On examination of the corresponding NH$_3$(2,2) optical depth map (Figure\,\ref{fig:OptDepth}), the feature is present with an optical depth of $\sim$0.2, otherwise the (2,2) optical depth is generally 30-50\% of that of the (1,1) optical depth throughout the main cloud component, where the (2,2) optical depth follows the same general trend as the (1,1) optical depth.

%Column density and temperature
\subsubsection{Temperature and the ortho-para-NH$_3$ ratio}\label{sec:ColdensTemp}
Figure\,\ref{fig:ColDens} displays the column densities of the (1,1), (2,2), (3,3), (4,4), (6,6) NH$_3$ states and the total NH$_3$ column density. The column density of each rotational state and the total NH$_3$ column density peaks towards the south-west region of the cloud and decreases steadily towards the north-east. This column density gradient may indicate a region of dense gas or shock-compression triggered by the W28 SNR, a scenario consistent with the detection of higher-energetic state \nh3 transitions. 

%temperatures
Towards the main cloud dense component, the calculated rotational temperatures (see Figure\,\ref{fig:OptDepth}) were relatively spatially constant ($T_{12}\sim$40-60$\pm$(10-40)\,K and T$_{36}\sim$260-295$\pm$(20-65)\,K). A non-LTE statistical equilibrium anlaysis of the same data (see below) yields a kinetic temperature consisitent with T$_{12}$, disfavouring T$_{36}$ as reliable measure of kinetic temperature. Within uncertainties, no spatial rotational temperature gradient is observed within either single rotational temperature map. The typical uncertainty was 10-20\% for T$_{36}$, but 30-60\% for T$_{12}$ towards the cloud. These errors lead to a hot component LTE column density with an error of 3-10\% and a cold component LTE column density with an error of 25-70\% (disregarding results for the component at the north-east corner, which has an error exceeding 100\%). After adding hot and cold components, the total column density had an error in the range 30-75\%. Percentage uncertainty maps for 7 key parameters calculated in this analysis are shown in Figure \ref{fig:Errors} and \ref{fig:Error2}.

%OPR
The ortho-para-NH$_3$ ratio (see Figure\,\ref{fig:OptDepth}) was observed to vary between 1 and 3.4, with a statistical uncertainty of $\sim$18-40\%. A high OPR ($>$2) is thought to indicate that the observed NH$_3$ originally formed on the surface of dust grains before being freed by heat or shock-collisions \citep{Umemoto:1999}. Upcoming work by \citet{deWilt:2016} will focus on NH$_3$ emission and the OPR values towards a population of gamma-ray sources, including W28-north, so we leave further investigation of this phenomenon as future work.

%\subsubsection{Non-LTE Model Cross-checks}\label{sec:NonLTE}
We performed non-LTE statistical equilibrium modelling to test the validity of our LTE analysis for 2 test regions, A and B, within the W28 NE cloud (see Figure\,\ref{fig:ColDens}). The RADEX statistical modeling software \citep{vanDerTak:2007} was employed to cycle through density and temperature parameter spaces to retrieve the best-fit (single component) values consistent with measured NH$_3$ inversion line observations via a $\chi ^2$-minimisation method. Parameter solutions were found for ortho and para-NH$_3$ emission lines in joint (ortho$+$para lines) analyses, using observed column density constraints, line intensities and NH$_3$(1,1) optical depths as inputs. An OPR of 2.3$\pm$0.5 was imposed for Region\,B, consistent with observations, but the OPR was assumed to be unity for Region\,A, where the OPR could not be calculated. The density-space between $10$\,cm$^{-3}$ and $10^{10}$\,cm$^{-3}$, and the temperature-space between 10 and 400\,K were tested. Molecular H$_2$ was assumed to be the only collision partner in our simulations.

NH$_3$ non-LTE analyses yielded temperature solutions consistent with T$_{12}$ in a $\chi ^2$ minimisation process for both Regions A and B, with temperatures of $\sim$55-70 and $\sim$35\,K, respectively. These analyses also suggest that the emitting clumps within Region\,A have a density of $\sim 2\times 10^5$-$2\times 10^6$\,cm$^{-3}$. Non-LTE analyses also yielded a degenerate solution for the density of Region\,B of $\sim 10^4$ and $ > 10^8$\,cm$^{-3}$, perhaps demonstrating a limit of single-component non-LTE modelling for this region. The former solution is the approximate critical density of NH$_3$(1,1), so is consiedered to be the most likely solution for Region\,B. This density information is revisited in the following section.

\subsection{Column density, Filling Factor and NH$_3$ Abundance}
\label{sec:massdens}
In \citet{nicholas2011a}, the dense gas component of the NE cloud was investigated using \nhone emission data less sensitive to those used here. The mass and density were calculated to be 1600\,\msun and $\sim800$\,\cmcube respectively, and further radiative transfer modelling with MOLLIE software suggested that the NE cloud had a mass $>1300$\,\msun, however this value was calculated under the assumption that [$\nh3$]/[H$_2$]$\sim2\times10^{-8}$. In this paper, the variation in Galactic NH$_3$ abundance is considered instead, and the reverse process is employed, i.e. a mass is used to calculate the abundance of Ammonia in the W28 NE cloud.

%can vary in an environment with high ionisation rate, like that observed towards the W28 north-east cloud \cite{Bayet:2011}, so the previous NH$_3$-derived mass may be incorrect. In this paper, we instead estimate the NH$_3$ abundance.

%(assuming [\nh3]/[H$_2$]$\sim2\times10^{-8}$). Here, we present an improved estimate of the mass and density of the NE cloud using our 12\,mm mapping data with improved sensitivity and the PbP analysis technique.

Observations of the CS(1-0) transition, which has a similar critical density to NH$_3$(1,1), found that the mass of the dense gas component of the NE cloud is $\sim$5.6$\times10^{4}$\,\msun \citep{Nicholas:2012}, in agreement with previous CO-derived mass estimates \citep{hess:w28}. CS(1-0) emission covers a region 30\% larger than the area that passed the quality checks in this NH$_3$ analysis.
%($\sim$1.0$\times$10$^{38}$\,cm$^2$). Accounting for this area-difference, CS(1-0) emission implies that $\sim$1.3$\times$10$^{61}$ H$_2$ molecules are within the NH$_3$-traced region highlighted in Figures\,\ref{fig:OptDepth}-\ref{fig:ColDens}, \textbf{making an average H$_2$ column density of $\sim$10$^{23}$\,cm$^{-2}$. 
The average CS-derived H$_2$ column density towards the NH$_3$-traced region highlighted in Figures\,\ref{fig:OptDepth}-\ref{fig:ColDens} is $\sim$10$^{23}$\,cm$^{-2}$. This value is supported by the implied optical extinction of $\sim$50 \citep[e.g.][]{Guver:2009}, which is consistent with the extinction value of 57 derived from infrared emission\footnote{http://irsa.ipac.caltech.edu/applications/DUST/} using a visual extinction to reddening ratio of 3.1 \citep{Schlafly:2011}. The average NH$_3$ column density within the region is (1.5$\pm$0.6)$\times$10$^{14}$\,cm$^{-2}$, leading to an estimated $\nh3$ abundance of [$\nh3$]/[H$_2$]$\sim (1.2 \pm 0.5)\times10^{-9}$. 

NH$_3$ abundance is observed to vary in different Galactic environments, with NH$_3$ molecules being released from dust-grains in warm environments, while being vulnerable to photo-dissociation in ionising UV and CR radiation fields. Typical NH$_3$ abundance values inside infrared-dark clouds are of the order 10$^{-8}$ \citep{Flower:2004,stahler2005,Rizzo:2014}, with hot ($>$100\,K) regions sublimating enough NH$_3$ from dust grain mantles (e.g \citealt{tafalla2004}) to make the gas-phase abundance increase, sometimes as high as $\sim$10$^{-6}$ \citep{Osorio:2009}. Such behaviour is also observed towards some shocks to various levels, e.g. [$\nh3$]/[H$_2$]$\sim$10$^{-6}$ in the bipolar outflow L1152 \citep{Tafalla:1995} and $\sim$10$^{-8}$ in the Wolf-Rayet nebula NGC\,2359 \citep{Rizzo:2001}. Photo-dissociation can have the opposite effect on NH$_3$ abundance, like near the intense UV field of the luminous blue variable star G79.29+0.476 \citep{Rizzo:2014}. Cosmic ray (CR) dissociation can also play a role according to modelling of the chemistry in CR-dominated regions, with ionisation rates above 10$^{-16}$\,s$^{-1}$ resulting in a significant decrease in NH$_3$ abundance \citep{Bayet:2011}. Although the average ionisation rate inside dense clouds might be as low as 10$^{-18}$-10$^{-17}$\,s$^{-1}$ \citep[e.g.][]{Padovani:2015}, direct measurements in the far north of the NE cloud suggest ionisation rates on the order of 10$^{-15}$\,s$^{-1}$ \citep{Vaupre:2014}.

The W28 NE cloud is likely a clumpy region and this can be investigated by by estimating the filling factor, $f$. By assuming a spherical clump of density, $n_{H2}$, and line-of-sight thickness, $L$, within the Mopra beam area, the column density can be expressed as $N_{H2}=f.n_{H2}.L$. The filling factor in this case would be the ratio of the cross-sectional area of the emitting clump, $\pi.(L/2)^2$, and the beam area, $\pi.(D/2)^2$, leading to $f=\left( L/D \right)^2$. Combining the column density and filling factor equations then allows an estimation of the filling factor,
\begin{equation}
\label{equ:FillFac}
f=\left(\frac{N_{H2}}{n_{H2}.D}\right)^{\frac{2}{3}}
 ~=~\left(\frac{N_{NH3}}{\chi . n_{H2}.D}\right)^{\frac{2}{3}}
\end{equation}where $\chi$ is the NH$_3$ abundance with respect to molecular hydrogen. D$\sim$1.2\,pc corresponds to the Mopra beam FWHM at 2\,kpc. From the non-LTE analyses, Regions A and B were estimated to have densities of $\sim 2\times 10^5$-$2\times 10^6$ and $\sim 10^4$\,cm$^{-3}$, 
%(with a degenerate solution of $>$10$^8$\,cm$^{-3}$)
respectively (see Section\,\ref{sec:ColdensTemp}). Inserting these values into Equation\,\ref{equ:FillFac} yields filling factors of $\sim 0.05$-$0.25$ and $\sim$2 for Regions A and B, respectively 
%($< 4\times 10^{-3}$ for degenerate density solution of $>$10$^8$\,cm$^{-3}$ for B). 
These estimates suggest that emitting NH$3$ clumps are distributed on scales from 0.2$^{\prime}$ to scales comparable to the beam FWHM (2$^{\prime}$). Interferometric NH$_3$ observations of finer angular resolution may be able to resolve such structure within the Mopra beam. Indeed, SiO(1-0) observations already show features on a 1$^{\prime}$-scales (see Figure\,\ref{fig:comparison1} and \citealt{Nicholas:2012}).
%( 10^23 / (10^6 * 10 *3e-19) )^(2/3)

Given the complexity of this region, the estimated NH$_3$ abundance of $(1.2 \pm 0.5)\times10^{-9}$ from the LTE analysis presented in this paper may be the result of a balancing between NH$_3$-release from dust grains and NH$_3$-destruction pathways resulting from radiation and CRs. An elevated OPR, like that observed in Region\,B, suggests that NH$_3$ has been released from dust grains \citep{Umemoto:1999}. On the other hand, there is no indication that an accompanying NH$_3$ abundance increase has occurred. In fact, the NH$_3$ abundance may be an order of magnitude below that of a typical infrared-dark cloud, despite the high temperature and shocked environment (see Section\,\ref{ssec:veldisp}). This low abundance may suggest that an NH$_3$ destruction mechanism is playing a significant role. Through this line of reasoning, a high-OPR/low-abundance combination may indeed be another piece of evidence to show that the NE cloud is heavily influenced by the W28 SNR, both directly through shock collision and shock heating, and indirectly from an enhanced ionisation rate. Such a scenario would require modelling of gas-phase NH$_3$ production and destruction beyond the scope of this work. We note that in our calculations, the NH$_3$ abundance is proportional to the CS abundance assumed by \citet{Nicholas:2012}, thus an alternative interpretation for our data towards this cloud is that the CS abundance is enhanced by a factor of $\sim$10 in the region, while the NH$_3$ abundance remains average ($\sim 2\times10^{-8}$). Certainly CS abundance can fluctuate due to freeze-out onto grains \citep[e.g.][]{tafalla2004}.

\section{Summary, Conclusions and Future Work}
\label{sec:conc}
We reported on deep mapping observations towards the shocked molecular cloud north-east of W28, with a focus on detecting multiple \nh3 inversion transitions. % to calculate gas parameters on a pixel-by-pixel scale. 
The NE cloud has a remarkable spatial match with the gamma-ray source HESS J1801-233, so constraints on the mass distribution are important for hadronic gamma-ray production models of the region, while the observed chemistry serves as an observational constraint on CR ionisation and propagation. Spectral line observations are steps towards parameter constraints associated with the NE cloud of W28.

These observations revealed that the dense component of the NE cloud is much more extended than previously reported. This is the case for all the detected inversion transitions. Towards the cloud, strong \nhthree, NH$_3$(4,4) and NH$_3$(6,6) emission suggest this is a region of high gas temperature. Furthermore, new evidence for shocked gas is provided by \nhone\,-\nhthree velocity dispersion maps that resolve a new \nh3 component on the W28 side of the NE cloud. %This highlights the usefulness of PbP analysis in mapping out gas parameters and revealing internal dynamics of molecular clouds.

Gas parameter maps were derived from NH$_3$ emission via a method that assumes Local Thermodynamic Equilibrium (LTE) and were checked against non-LTE statistical equilibrium models. NH$_3$ column densities on the order of 10$^{14}$\,cm$^{-2}$ and temperatures in the range 35-60\,K were observed within the NE cloud of W28.

The ortho-para-NH$_3$ ratio (OPR) was investigated, revealing a subregion with an elevated OPR ($>$2), characteristic of regions where NH$_3$ is liberated from dust-grain mantles. Comparing our measurements with a previously-published CS-derived mass estimate, no corresponding NH$_3$ abundance enhancement was observed ([\nh3]/[H$_2$]$\sim (1.2 \pm 0.5)\times10^{-9}$), possibly suggesting the existence of an NH$_3$ destruction mechanism. More detailed modelling of gas-phase NH$_3$ production and destruction may be required to investigate this result.

Future work to improve the angular resolution and sensitivity of TeV gamma-ray images will allow a detailed comparison of the gamma-ray emission and cosmic ray target material (the gas), while considering the time-dependent effect cosmic ray propagation may also allow the analysis of features in the GeV to TeV gamma-ray spectrum towards SNRs (e.g. \citealp{Gabici:2007, Maxted:2012}).

\section*{Acknowledgements}
This work was supported by Australian Research Council grants (DP0662810, DP1096533). 
The Mopra Telescope is part of the Australia Telescope and is funded by the Commonwealth of Australia for operation as a National Facility managed by CSIRO. The University of New South Wales Mopra Spectrometer Digital Filter Bank used for these Mopra observations was provided with support from the Australian Research Council, together with the University of New South Wales, University of Sydney, Monash University and the CSIRO.

We would like to thank the anonymous referee whose comments served to increase the quality of our manuscript and maximise the exploitation of our data.

%\begin{figure*}
%\includegraphics[height=0.49\textheight]{./figures/PVplotMultipanel.eps}
%\caption{RA vs. \vlsr position velocity (PV) plots for the \nhone, \nhtwo, \nhthree \& \nh3(4,4) emission. The apparent direction of the W28 shock is indicated in the top-left image.}
%\label{fig:pvplots123}
%\end{figure*}

\appendix

\section{Uncertainties of key parameters}
\begin{figure*}
\includegraphics[width=0.99\textwidth]{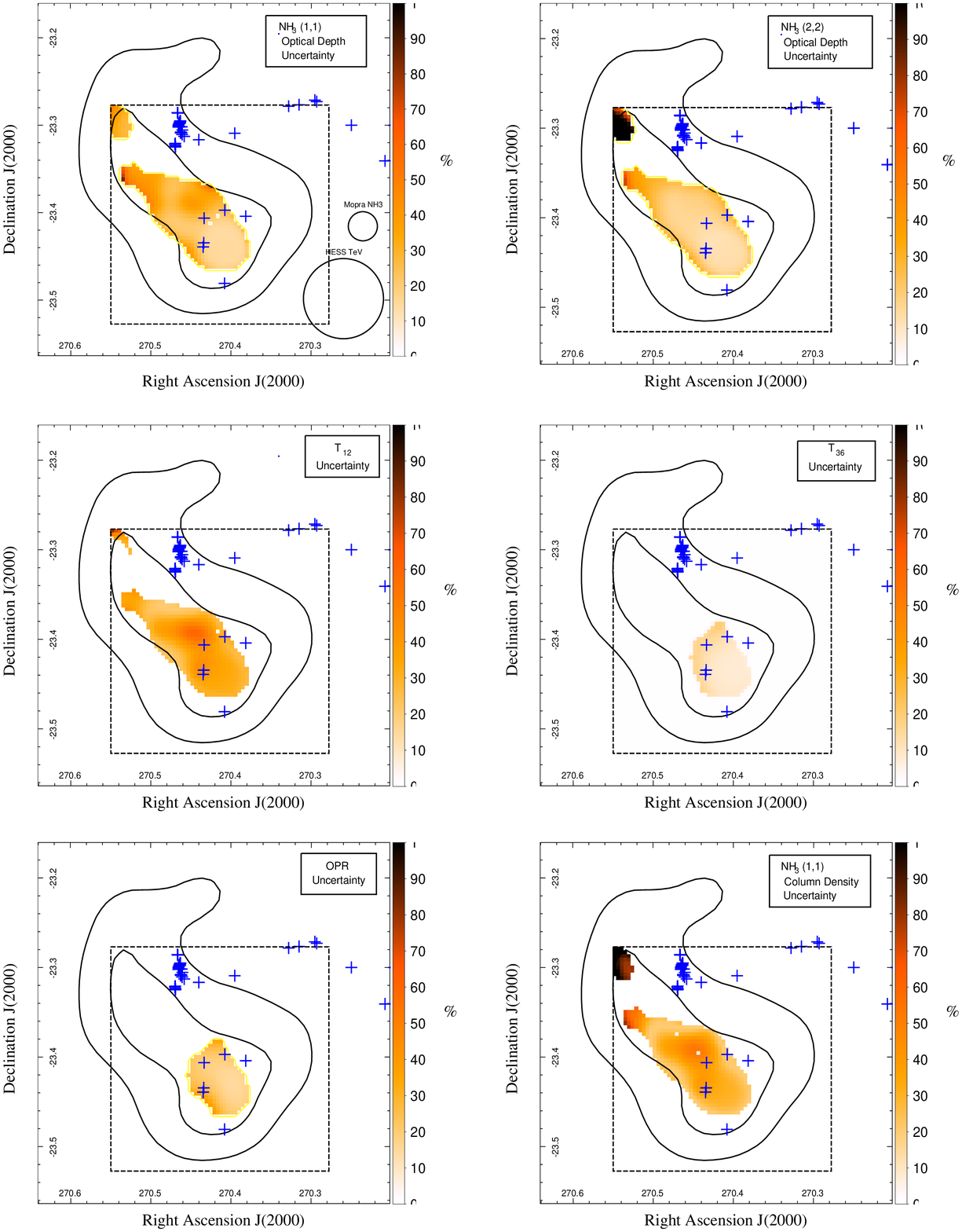} 
\caption{Images of the uncertainty in NH$_3$(1,1) and (2,2) optical depth (top left and right, respectively), rotational temperatures (middle, left and right), the OPR (bottom left) and NH$_3$(1,1) column density (bottom, right). OH masers are represented by blue crosses in all maps.}
\label{fig:Errors}
\end{figure*}

\begin{figure}
\includegraphics[width=0.495\textwidth]{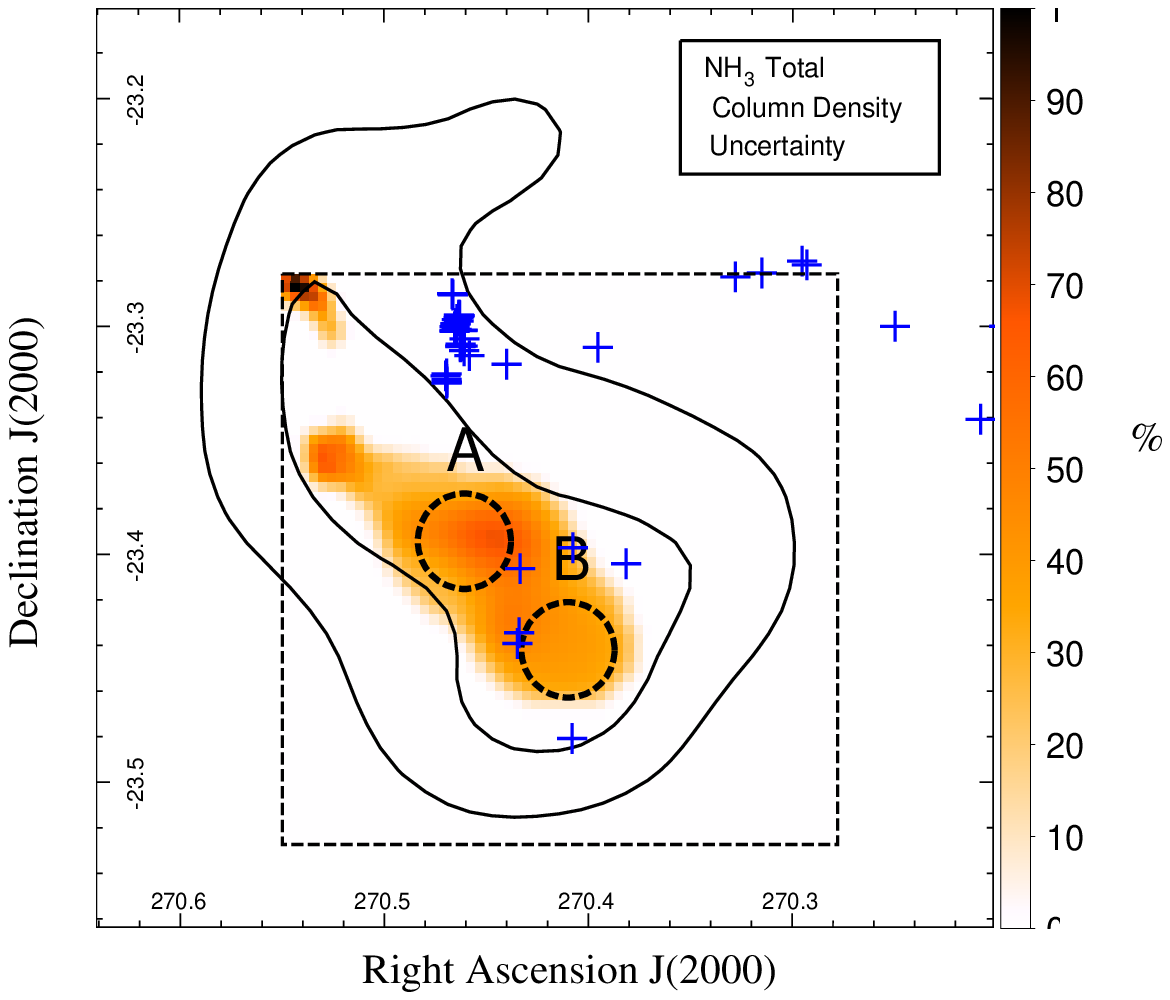}
\caption{Image of the uncertainty of the total NH$_3$ column density. OH masers are represented by blue crosses and Regions\,A and B, which were used for non-LTE cross checks (see Section\,\ref{sec:ColdensTemp}) are displayed.}
\label{fig:Error2}
\end{figure}

\end{document}